\documentstyle[preprint,aps]{revtex}
\input{epsf}

\begin{document}
\draft
\title{Statistics of energy spectra of a strongly disordered system of
interacting electrons}

\author
{R. Berkovits$^1$ and B. I. Shklovskii$^2$}

\address{$^1$
The Minerva Center for the Physics of Mesoscopics, Fractals and Neural
Networks,\\ Department of Physics, Bar-Ilan University,
Ramat-Gan 52900, Israel}

\address{ $^2$
Theoretical Physics Institute, Department of Physics, University of
Minnesota,\\ Minneapolis MN 55455}

\date{\today}
\maketitle

\begin{abstract}

Statistics of many particle energy levels of a finite two-dimensional
system of interacting electrons
is numerically studied. It is shown that the statistics of these levels
undergoes a Poisson to Wigner crossover as the strength of the disorder is
decreased. This transition occurs
at a similar strength of disorder as the one-electron delocalization crossover
in a finite 2d system and develops
almost simultaneously at all energies. We interpret this crossover in terms
of delocalization in the space of occupation numbers of strongly bound and
compact
electron-hole pairs (excitons).

\end{abstract}


\narrowtext

\newpage

\section{Introduction}

The statistics of the quantum energy spectra of a disordered system of
noninteracting electrons, for example, the Anderson model,
was shown to be a good diagnostic tool to study an insulator-metal
transition\cite{sssls}.
It was discovered that the crossover from Poisson distribution
of the nearest neighbor level spacings to
a Wigner distribution sharpens with the
system size. Finite size scaling then permits to find out
whether the transition exists and if it does to calculate
quite accurately transition point and indexes \cite{sssls,percol}.

Recently attention has started to shift in the direction of spectral statistics
of the total energy of a finite disordered system of interacting
electrons \cite{agkl,js,sil,gs,wpi,ba}.
Good examples of such systems are quantum dots
(here we are talking about
the energies of the excited states of the dot
 and not about the charging spectrum). All
previous works that we know of deal with metallic systems which are well above
insulator-metal transition. For such system a Poisson-Wigner (P-W)
crossover was predicted with a growing energy or interaction
strength \cite{agkl,js,sil,gs,ba}.
The relation of this
crossover to delocalization in Fock space and the decay of one electron
states was discussed \cite{agkl}.
The intriguing question of whether such statistics
can work as a tool to study
insulator-metal transition has not yet been addressed.
In this paper we concentrate on the two-dimensional case where the
very existence of a
transition in an interacting system is under debate for a long time.

We started this work by
exploring whether we can obtain any evidence for the transition by studying
the finite size effects in the statistics
of levels of the total energy of a many particle system.
As far as we know this is the first study
of such statistics
at the insulating side of the
delocalization crossover and at the crossover itself.
We did not find any
conclusive evidence for this crossover.
Instead we found that the
delocalization crossover as function of the  decreasing disorder
generates a P-W crossover in the excited state statistics
which takes place almost uniformly at
all energies larger than the single level spacing
(the energy of the many body excited states are calculated from the many body
ground state).
We interpret the excited states with high energies as
consisting of several electron-hole excitations. Each electron-hole pair
is bound by Coulomb interaction. We  call such excitations excitons.
At strong enough disorder states are localized in the
space of states with different number of excitons or,
in other words, states of very close energies but with different number
of excitons do not mix coherently. As a result, the  nearest level distribution
function of many body states is Poissonian.
With decreasing disorder the rate of decay of an
exciton to smaller energy excitons becomes of the order of spacing
between many particle levels. The fact that
this P-W crossover is almost independent on energy and interaction
differs drastically from what happens in metallic
samples\cite{agkl,js,sil,gs,ba}. This occurs because the
density of states of the excitons and the exciton-exciton
interactions are different from the
ones for weakly interacting electron-hole pairs in a metal.

\section{Model}

The numerical study is based on the following interacting many-particle
tight-binding Hamiltonian:
\begin{eqnarray}
H= \sum_{k,j} \epsilon_{k,j} a_{k,j}^{\dag} a_{k,j} - V \sum_{k,j}
(a_{k,j+1}^{\dag} a_{k,j} + a_{k+1,j}^{\dag} a_{k,j}) + h.c
+ H_{int},
\label{hamil0}
\end{eqnarray}
where
$\epsilon_{k,j}$ is the energy of a site ($k,j$), chosen
randomly between $-W/2$ and $W/2$ with uniform probability, and $V$
is a constant hopping matrix element.
The interaction Hamiltonian is given by:
\begin{equation}
H_{int} = U
\sum_{k,j>l,p} {{a_{k,j}^{\dag} a_{k,j}
a_{l,p}^{\dag} a_{l,p}} \over
{|\vec r_{k,j} - \vec r_{l,p}|/b}}
\label{hamil2}
\end{equation}
where $U=e^2/b$ and
$b$ is the lattice unit.

We consider $3 \times 3$, $4 \times 3$, and $4 \times 4$
dots with $m=9,12,16$ sites and $n=3,4,4$
electrons. The $M \times M$ Hamiltonian (where $M=(^m_n)$)
matrix is numerically
diagonalized and all the eigenvectors $|\Psi_j\rangle$ and eigenvalues
$E_j$ are obtained.
The strength $U$ of the interaction is varied between $0-30V$.
and the disorder strength is chosen to be between $W=5V$ and $W=100V$.
Usually results are averaged over $1000$ realizations.

We will use the energy level statistics as an indication
of the Anderson transition
in Fock space. A convenient way to characterize the change
in the statistics of a system proposed
in Ref. \cite{sssls} is to study the parameter $\gamma$
defined as
\begin{equation}
\gamma= {{\int_2^\infty P(s) ds - e^{-\pi}}\over{e^{-2}- e^{-\pi}}},
\label{gamma}
\end{equation}
where $P(s)$ is the distribution of the normalized level spacings
$s=E_j -E_{j-1}/\langle E_j -E_{j-1} \rangle$,
where $\langle \ldots \rangle$ denotes an average over different
realizations of disorder. For an infinite system $\gamma$
changes sharply from $\gamma=1$ in the localized regime to $\gamma=0$
in the extended regime. For a finite system
the change is gradual.

\section{Results and Discussion}

\begin{figure}
\centerline{\epsfxsize = 4in \epsffile{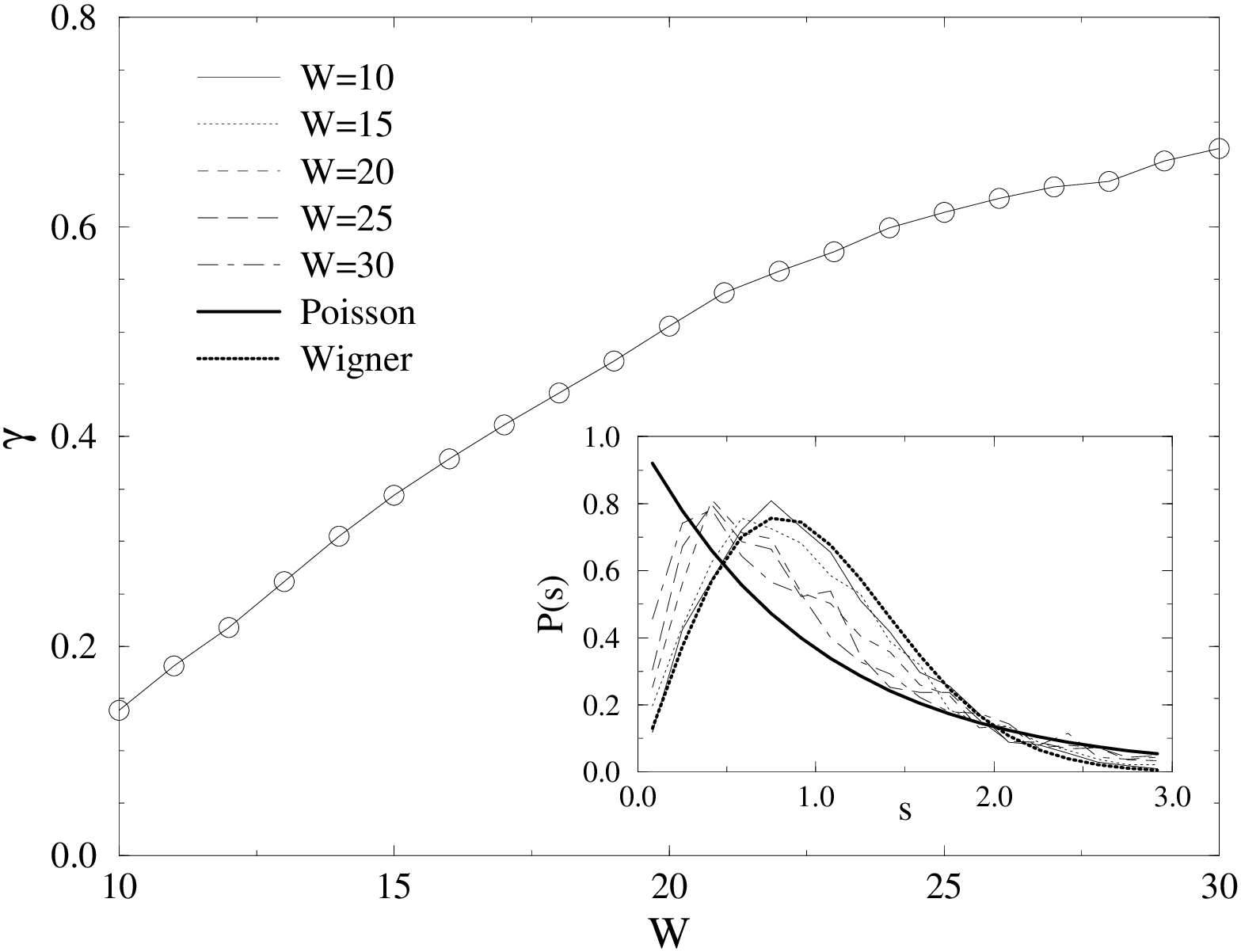}}
\caption{The values of $\gamma$ for the
spacing between the many particle ground-state and the first excited state
as function of the disorder
$W$ for a  $4 \times 3$ lattice with $4$ electrons. Inset: the distribution
$P(s)$ for different values of disorder and a given interaction ($U=10V$)}
\label{fig1}
\end{figure}

We begin by presenting the behavior of $\gamma$ for the non-interacting
spacing between the many particle ground state and the first excited state.
A crossover from a Wigner to Poisson
statistics as $W$ increases is clearly seen (Fig. \ref{fig1}).

From finite size
studies of one electron problem
one can estimate that the point for which the localization
length $\xi$ is of the system size corresponds to
$\gamma=0.35$ \cite{sssls}. In our case such
values of $\gamma$ are realized at $W \sim 15V$.
This does not contradict to the calculation by MacKinnon and
Kramer of $\xi(W=15)=2.2b$ \cite{mk}. So the P-W crossover
for the first excitation energy happens close to
the finite size delocalization crossover
in noninteracting system, which is expected.

\begin{figure}
\centerline{\epsfxsize = 4in \epsffile{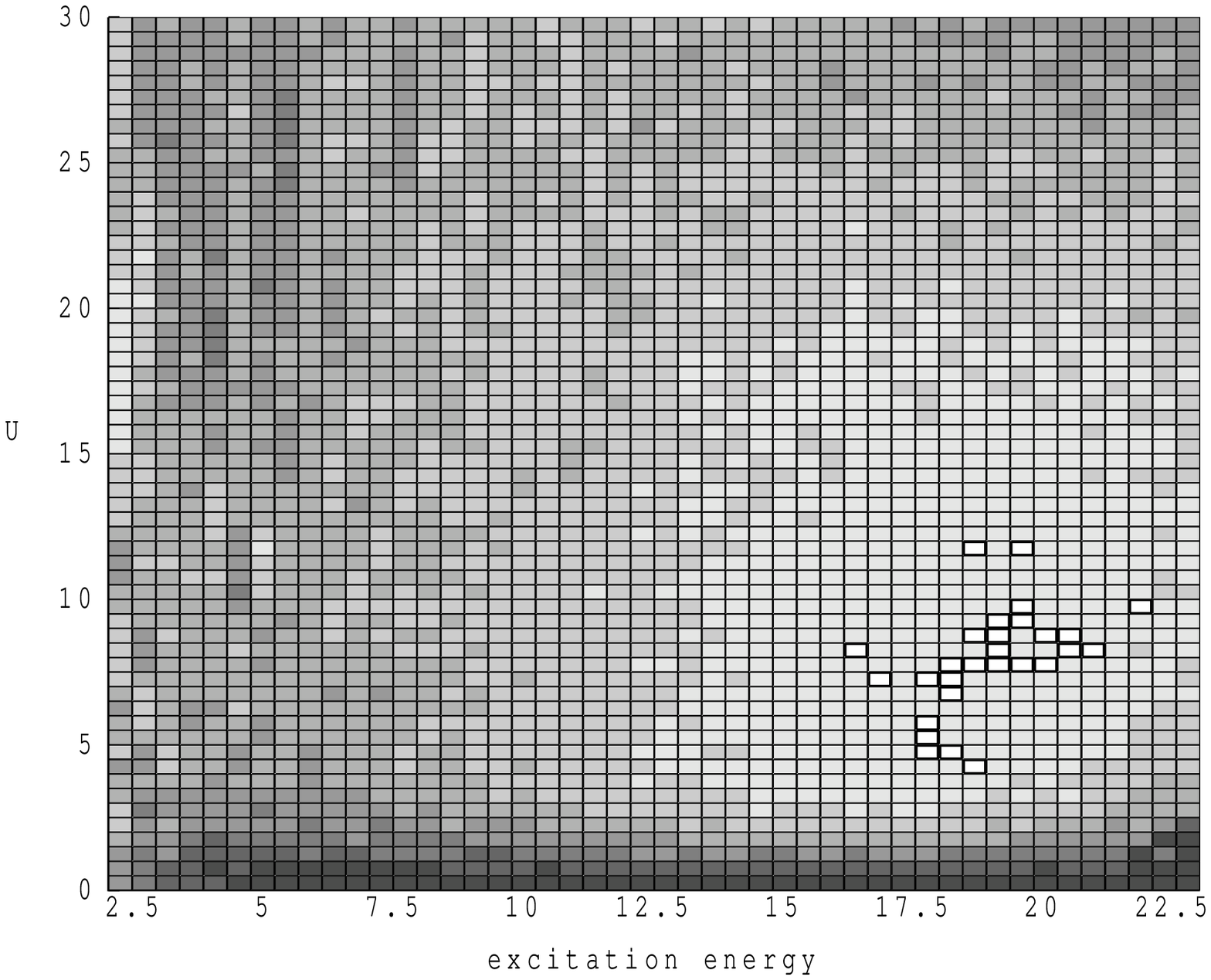}}
\leftline{(a)}
\centerline{\epsfxsize = 4in \epsffile{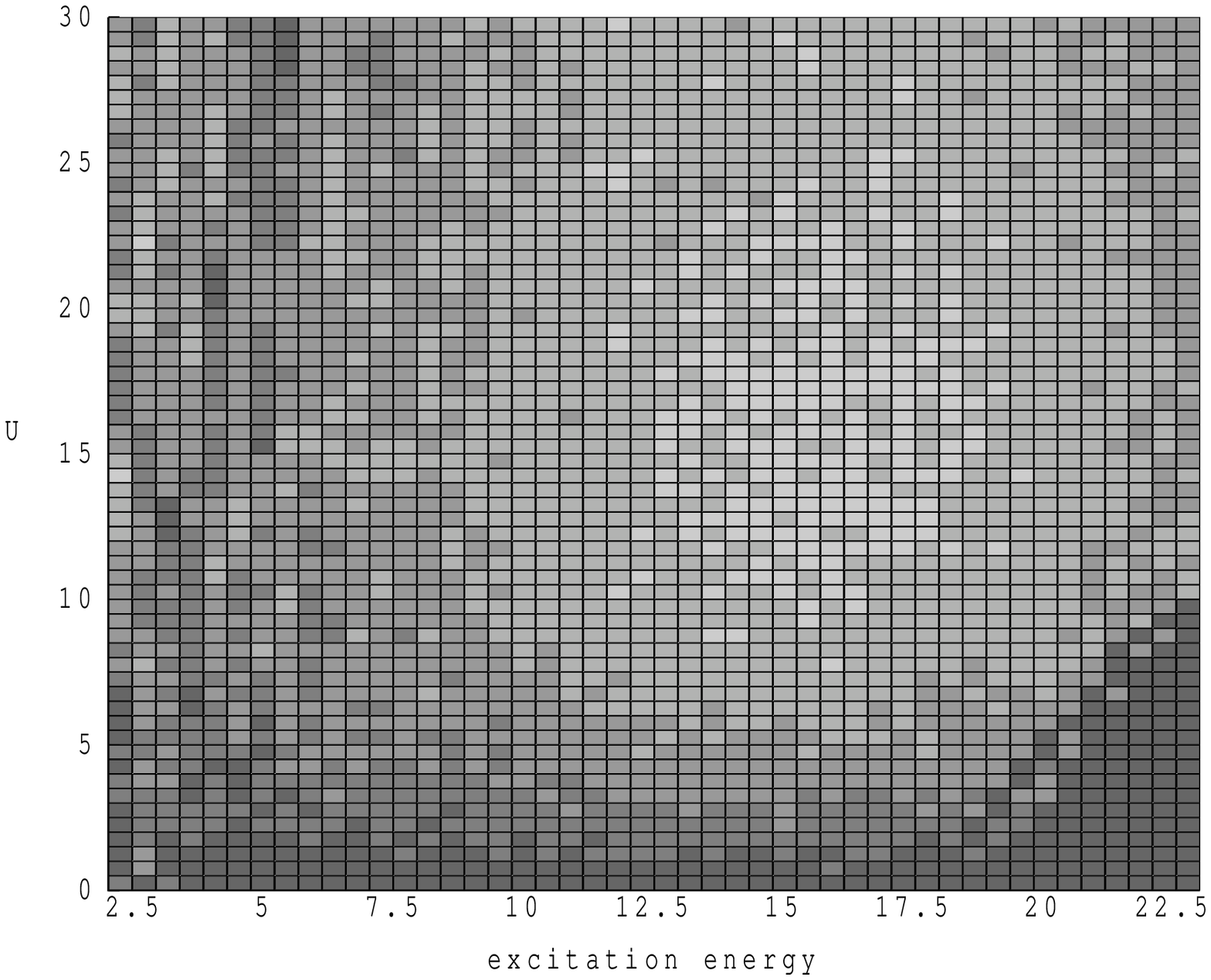}}
\leftline{(b)}
\centerline{\epsfxsize = 4in \epsffile{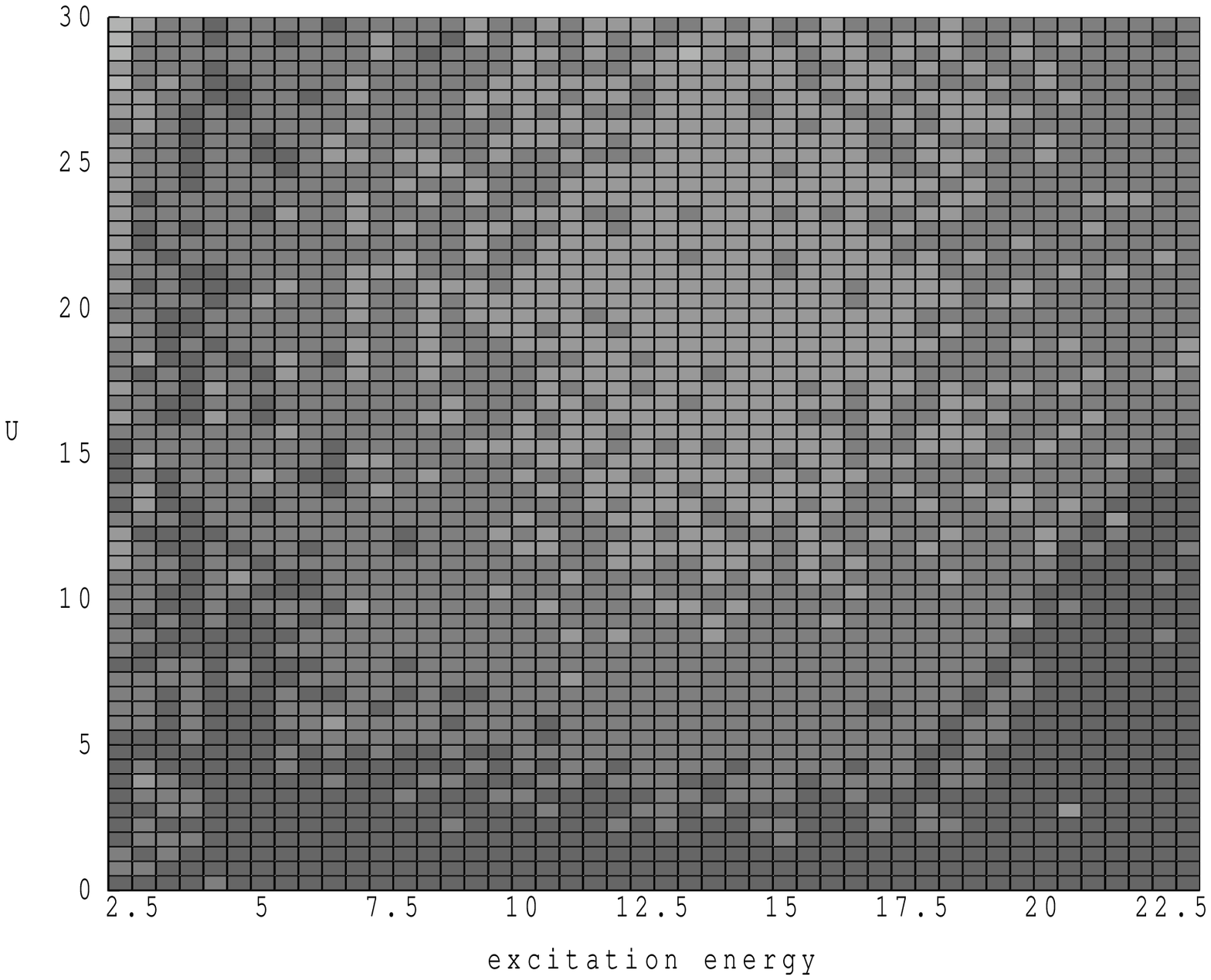}}
\leftline{(c)}
\centerline{\epsfxsize = 4in \epsffile{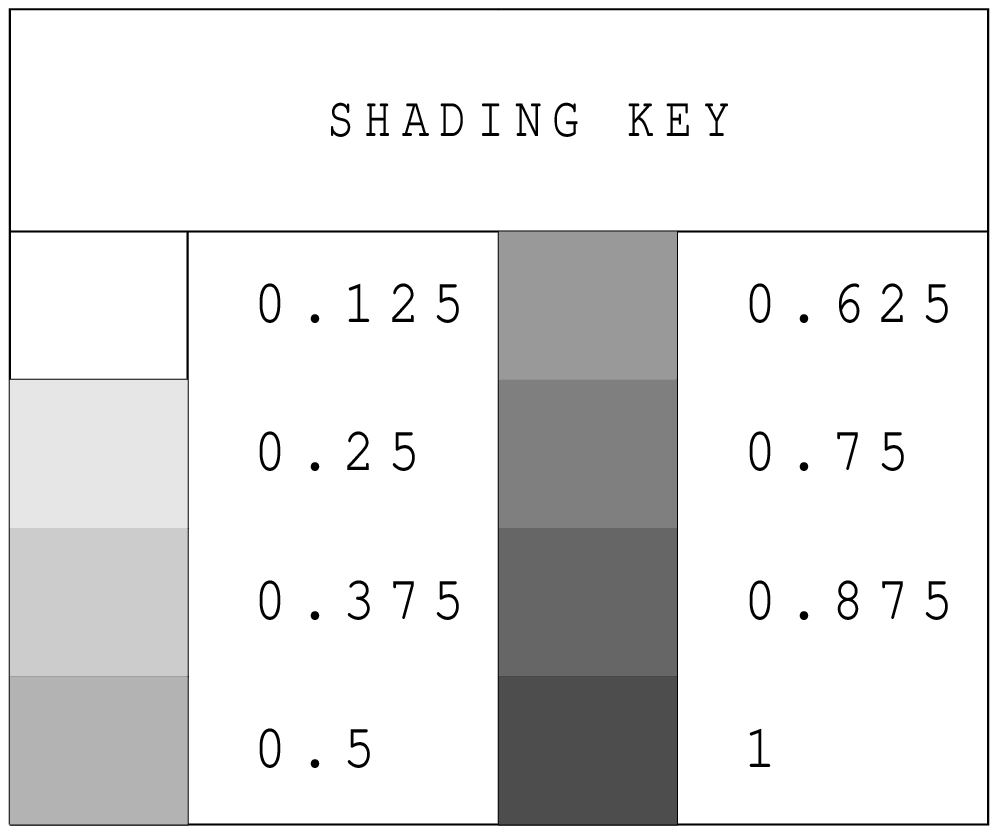}}
\caption{A gray scale map of
the values of $\protect \gamma$ as function of the interaction strength
$U$ and excitation energy in units of the single electron level spacing
$\Delta$
for a $4 \times 3$ lattice with $4$ electrons (a) $W=10V$,
(b) $W=20V$, (c) $W=30V$. The corresponding single level spacings are
(a)$\Delta=0.899V$,
(b)$\Delta=1.595V$, and (c)$\Delta=2.344V$ .}
\label{fig3}
\end{figure}

In the inset of Fig. \ref{fig1} we show the distribution of the level
spacings close to the many particle band center for $U=10V$.
A clear crossover from a
Wigner like behavior to a Poisson behavior can be seen. In order to
show this behavior for different excitation energies and interaction strength
we present in Fig. \ref{fig3}
gray scale maps of $\gamma$ for three different values of $W$.
The gray scale maps shows the average value
of $\gamma$ for the spacings between a many particle state with energy
$E$ above the ground state and the many particle state above it.
A general feature which appears with growing  disorder strength at
$W \sim 15$ is apparent - the statistics for energies $E>2 \Delta$ (where
$\Delta$ is the the single electron level spacing) is rather uniform
and does not depend strongly on interaction strength nor on excitation energy.
This feature becomes more pronounced as the disorder increases and
is in strong contrast to the situation in the metallic regime, in which
interesting features were seen as function of the interaction
strength and excitation energy.

This behavior clearly shows that at a large $W$ the different high energy
many particle states can be close in energy but nevertheless can have small
repulsion, i.e., interactions do not couple  different many particle states
no matter what energy is available. We interpret this behavior as the
result of the high energy many particle states being composed of several
electron-hole excitations (excitons). Neighboring many particle states
 usually are composed
of a different number of excitons and are related by
a very weak interaction matrix element between them
no matter how strong the interactions are. Thus, no repulsion
between the states  appears and the statistics is essentially
Poissonic for any interaction strength or energy.
The  P-W crossover as the disorder $W$ decreases is rather uniform and
shows no strong dependence on energy or interaction strength
(as long as one is still above the transition and
the interaction $U>2V$). We interpret
this crossover as result of the delocalization of the system in the space
of states
with different number of excitons which happens once the
matrix element for the decay
of a typical exciton into two smaller ones
becomes of the order of the spacing between many particle levels.

The energy independent P-W crossover revealed here differs drastically
from the predictions and calculations  made for Fock space delocalization
in metallic systems
\cite{agkl,js,sil,gs,wpi,ba}.
We relate this fact to the  difference between the excitons of the insulating
phase and the weakly interacting electron-hole pairs of the metallic samples.
Contrary to the latter ones, excitons consist of an
electron and a hole strongly bounded to each other by the Coulomb interaction.
In the limit of large $W$ and $U$ in the classical Coulomb glass
this exciton is just the classical compact
electron-hole pair excitation of Ref. \cite{es}. Due to
the existence of the Coulomb gap, electron-hole excitations corresponding
to the transfer of an
electron by a small distance (compact pair) are known to have a
constant density of states at small energies.
On the other hand, in a metallic dot
the joint density of states of weakly interacting electron-hole
pairs is linear in energy. We expect that such a difference is preserved
in the quantum system. To find the P-W crossover we have to use the density
of states of excitons and the matrix elements for an exciton decay.
We have already mentioned that there is a drastic difference between the exciton
density of states
of metallic and insulating samples. The matrix elements
of the exciton decay
should be different from the matrix element of the emission
of an electron-hole pair by a
free electron used in Refs. \cite{agkl,js,sil,gs,wpi} as well.

Thus a drastic difference between the P-W crossovers for metallic
and insulating cases seems to be natural. Unfortunately we could
not prove the uniform energy P-W crossover in the insulating case.

\begin{figure}
\centerline{\epsfxsize = 4in \epsffile{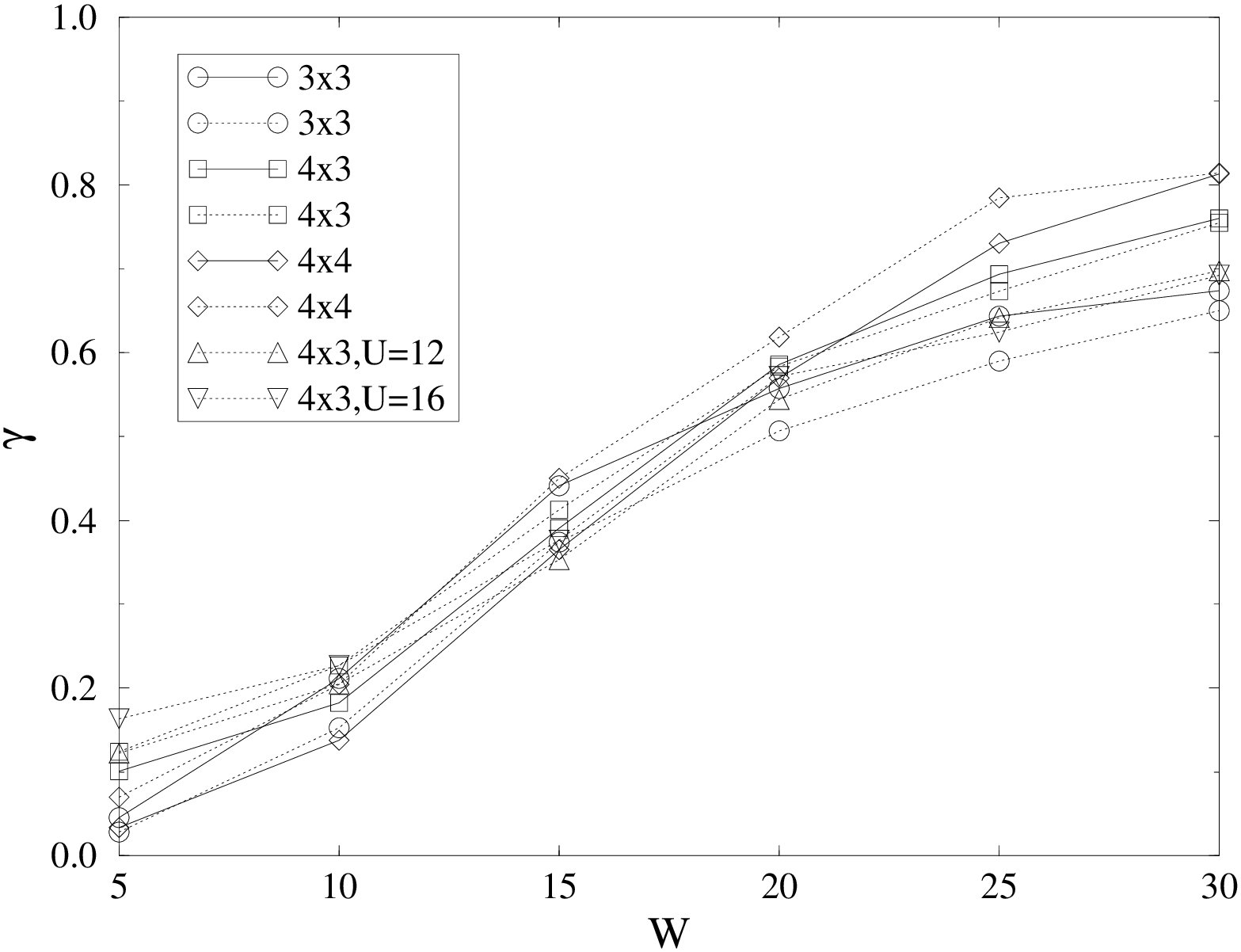}}
\caption{The values of $\gamma$ as function of disorder
$W$ for a $3 \times 3$ lattice with $3$ electrons,
$4 \times 3$ lattice with $4$ electrons,
$4 \times 4$ lattice with $4$ electrons,
at interaction strength $U=8V$, and for a $4 \times 3$ lattice with $4$
electrons also for $U=12V$ and $U=16V$.
The full line corresponds
to the value of $\gamma$ averaged over the lowest $3\%$ of the
spacings in the many particle spectrum, and the dotted line corresponds
to $\gamma$ averaged over the lowest $10\%$ of the spacings.}
\label{fig4}
\end{figure}

In Fig. \ref{fig4} we present a more quantitative description of
the P-W crossover. We show the results for $\gamma$ averaged over
$3\%$ and $10\%$ of the low lying many particle energy levels for
intermediate values of the interaction ($U=8V,12V,16V$) and for different
lattice sizes and electron numbers. It is obvious
that in all cases $\gamma$ increases as the disorder $W$ is enhanced.
This is a possible indication of the signature of the single electron
delocalization crossover on the many particle spectrum.
There is no significant difference between
the values of $\gamma$ for $3\%$ and $10\%$
of the spectrum, nor a strong dependence on interaction strength,
so that crossover indeed happens rather uniformly in energy and 
interaction strength.
As larger values of disorder are approached the difference becomes even smaller
and the many particle spectrum becomes even more uniform.

Because not much depends on energy there is an a priori
chance that this crossover somehow reflects an
insulator-metal transition
in a many particle interacting system.
As can be seen in Fig. \ref{fig4} there is no clear finite size behavior. This
probably means that we  are  dealing with a crossover, not a phase transition.

Before we conclude we want to comment on the
importance of excitons  introduced above. Now we concern
ourselves
with the localization of excitons in real space.
The following scenario seems likely (although we have not
found  a way to prove it using our numerical data).
Excitons are localized as long as the disorder
dominates and the many body level statistics is Poissonian.
There might exist a crossover regime where the
charge is still localized but the excitons become delocalized. In the case when
a metal-insulator transition exists (in three and possibly for the
interacting case in two dimensions) the
exciton delocalization happens on the insulating side of
the critical region of
the metal-insulator transition. This scenario can result in a
situation where electronic conductivity is exponentially small while the
electronic thermal conductivity changes as a power of temperature
~\cite{bur}.

Another consequence of the possible exciton delocalization is that
they can play a crucial role in low temperature variable range hopping. At low
temperatures they can assist electron hopping  much more
effectively than phonons. As a result the prefactor of
the variable range hopping can acquire an universal value $e^2/h$,
which was observed experimentally. This in turn
leads to a very simple microscopic interpretation of the dynamic scaling at
a number of quantum phase transition points,
such as the quantum Hall or
the superconductor-insulator transitions~\cite{PS,APS}.

Arguments for the
delocalization of two interacting electrons above the Fermi sea in a situation
where both of them are localized (as well
as the other electrons of the Fermi sea) were given in
Refs.~\cite{s,i}. In the case of electrons interacting via the
Coulomb interaction these arguments should not work
because the joint density of states of two electrons drastically decreases at
small energy due to the Coulomb gap in the one-electron density of states.
However, for a compact electron-hole pair, an exciton,  as we mentioned above,
Coulomb effects increase its density of states making these arguments
more plausible.
Actually these arguments were applied to the exciton before~\cite{i}, however,
the effect of the
Coulomb enhancement was not considered.

In conclusion, we found a P-W crossover in
the statistics of the
nearest neighbor spacings of many particle levels, which occurs almost
simultaneously
at all energies. To interpret this crossover we introduced
excitons and speculated that the crossover is related to the increase of
their interaction which in turn leads
to the transition from a description where
each state corresponds to a number of weakly interacting excitons to
new states which are delocalized in the space of the old ones. This
transition is similar to delocalization in Fock space in metallic samples
recently studied in Refs.~\cite{agkl,js,sil,gs,wpi,ba}.

\acknowledgements

We are grateful for  discussions to A. Yu. Dobin, A. A. Koulakov, M. M. Fogler
and D. L. Shepelyansky. R.~B. research was supported by The Israel Science
Foundations Centers of
Excellence Program.
B.~I.~S. acknowledges support from NSF under Grant
DMR-9616880.

\begin{references}

\bibitem{sssls} B. I. Shklovskii, B. Shapiro, B. R. Sears, P. Lambrianides,
and H. B. Shore, \prb {\bf47}, 11487 (1993).

\bibitem{percol} R. Berkovits and Y. Avishai, \prb {\bf 53}, R16125 (1996).

\bibitem{agkl} B. L. Altshuler, Y. Gefen, A. Kamanev and L. S. Levitov,
\prl {\bf 78}, 2803 (1997).

\bibitem{js} Ph. Jacquod and D. L. Shepelyansky, \prl {\bf 79}, 1837 (1997).

\bibitem{sil} P. G. Silvestrov ,\prl {\bf 79}, 3994 (1997).

\bibitem{gs} B. Georgeot and D. L. Shepelyansky, \prl {\bf 79}, 4365 (1997).

\bibitem{wpi} D. Weinmann, J. L. Pichard and Y. Imry ,
J. Phys. I (France) {\bf 7}, 1559 (1997).

\bibitem{ba} R. Berkovits and Y. Avishai, \prl {\bf 80}, 568 (1998).

\bibitem{mk} A. MacKinnon and B. Kramer, Z. Phys. B, {\bf 53}, 1 (1983).

\bibitem{es} B. I. Shklovskii, A. L. Efros, {\it Electronic properties
of doped semiconductors}, (Springer, Heidelberg (1984)).

\bibitem{bur}  A. L. Burin, L. A. Maksimov, and I. Ya. Polishchuk,
JETP Lett. 49, 785 (1989).

\bibitem{PS} D. G. Polyakov, B. I. Shklovskii, \prb {\bf 48}, 11167 (1993).

\bibitem{APS} I. L. Aleiner,  D. G. Polyakov,  B. I. Shklovskii,
 "Physics of    Semiconductors", World Scientific, p. 787, (1995)

\bibitem{s} D. L. Shepelyansky,\prl {\bf 73}, 2067 (1994).

\bibitem{i} J. Imry , Europhysics Lett. {\bf 30}, 405 (1995).


\end{references}
\end{document}